\documentclass[conference]{IEEEtran}
\IEEEoverridecommandlockouts

\usepackage[utf8]{inputenc} 
\usepackage[T1]{fontenc}    
\usepackage[hidelinks]{hyperref}
\usepackage{url}            
\usepackage{booktabs}       
\usepackage{amsfonts}       
\usepackage{nicefrac}       
\usepackage{microtype}      
\usepackage{lipsum}
\usepackage{graphicx}
\usepackage{multirow}
\graphicspath{ {./images/} }
\usepackage{xcolor}
\usepackage{amsmath}
\usepackage{subcaption}
\usepackage{cite}

\newtheorem{RQ}{Research Question}
\newtheorem{hypothesis}{Hypothesis}

\begin{document}

\title{Leveraging Code Cohesion Analysis to Identify Source Code Supply Chain Attacks}

\author{
Maor Reuben$^{1}$,
Ido Mendel$^{1}$,
Or Feldman$^{1}$,
Moshe Kravchik$^{1}$,
Jacob Rimer$^{2}$,
Mordehai Guri$^{1}$,
and Rami Puzis$^{1}$\\
\IEEEauthorblockA{
$^{1}$\textit{Ben-Gurion University of the Negev}, Beer Sheva, Israel \\
$^{2}$\textit{Independent Researcher} \\
\{maorreu, idoman, orfel, moshekr, gurim\}@post.bgu.ac.il, jacobrmft@gmail.com, puzis@bgu.ac.il
}
}

\maketitle

\begin{abstract}
Supply chain attacks significantly threaten software security with malicious code injections within legitimate projects. 
Such attacks are very rare but may have a devastating impact. 
Detecting spurious code injections using automated tools is further complicated as it often requires deciphering the intention of both the inserted code and its context.  
In this study, we propose an unsupervised approach for highlighting spurious code injections by quantifying cohesion disruptions in the source code. 
Using a name-prediction-based cohesion (NPC) metric, we analyze how function cohesion changes when malicious code is introduced compared to natural cohesion fluctuations. 
An analysis of 54,707 functions over 369 open-source C++ repositories reveals that code injection reduces cohesion and shifts naming patterns toward shorter, less descriptive names compared to genuine function updates. 
Considering the sporadic nature of real supply-chain attacks, we evaluate the proposed method with extreme test-set imbalance and show that monitoring high-cohesion functions with NPC can effectively detect functions with injected code, achieving a Precision@100 of 36.41\% at a 1:1,000 ratio and 12.47\% at 1:10,000.
These results suggest that automated cohesion measurements, in general, and name-prediction-based cohesion, in particular, may help identify supply chain attacks, improving source code integrity.
\end{abstract}


\section{Introduction}
\label{sec:intorduction}
In today's software-driven world, ensuring the security and reliability of software systems has become a critical concern. 
The increasing prevalence of supply-chain compromises, where malicious actors exploit vulnerabilities in the source code supply chain to introduce unauthorized modifications, poses significant threats to the integrity and trustworthiness of software applications~\cite{okafor2022sok}. 

A prominent example is the SolarWinds incident in 2020, where attackers embedded a backdoor in an update for the Orion platform, granting unauthorized remote access to thousands of corporate and government systems and causing extensive data breaches~\cite{sterle2021solarwinds}. 
Similarly, Kaseya's software was compromised by ransomware, resulting in a \$70 million extortion demand~\footnote{\url{https://www.kaseya.com/}}~\cite{robinson2022new}.

Traditional methods for detecting these attacks primarily rely on rule-based approaches~\cite{vu2020fork,Microsoft2019OSS} and machine learning- driven static or dynamic code analysis techniques~\cite{duan2020towards,garrett2019detecting}. 
However, these techniques depend heavily on known malware signatures, behavior patterns, or supervised learning models that require labeled datasets. 
As attackers continue to adapt, employing sophisticated anti-analysis techniques to evade detection~\cite{guo2023empirical}, there is a pressing need for unsupervised approaches that can identify subtle anomalies without prior knowledge of malicious patterns.

One promising avenue for addressing this challenge is the detection of code cohesion anomalies. 
Code cohesion, a fundamental software quality attribute, measures the extent to which elements within a module (e.g., functions or classes) collaborate to achieve a common objective~\cite{rosenberg1997software}. 
High cohesion typically indicates robust, maintainable, and readable code, whereas low cohesion can signal the introduction of unrelated responsibilities.
By continuously monitoring code cohesion, deviations from expected levels can serve as early warning signs of both security threats and maintainability concerns.

In this paper, we introduce \textbf{an unsupervised methodology for detecting potential supply chain attacks by leveraging pre-trained language models (PLMs) to assess code cohesion.} 
Our approach frames cohesion estimation as a function purpose prediction problem, where the alignment between a function’s body and its intended name serves as a proxy for cohesion. 
Core to our approach is the idea that the cohesion of a function is determined by how different elements in the code contribute to its purpose \cite{bieman1995cohesion}.
By evaluating functions for deviations from normal cohesion levels, our method detects possible compromises without relying on labeled data or prior knowledge of malicious code.
Additionally, our approach supports software maintainers by identifying functions that may require refactoring due to cohesion degradation, assisting in long-term software evolution and quality assurance efforts.

We evaluate our approach using a dataset of 369 open-source C++ projects, capturing comprehensive version histories for functions in each project (Section~\ref{sec:dataset}). 
Through simulated malware injection scenarios, we demonstrate that our method effectively differentiates between legitimate maintenance changes and malicious code alterations.
Moreover, we assess how our technique can aid in identifying unintended cohesion drift over time, making it a valuable tool for both security enforcement and software maintenance practices.

The main contributions of this study are threefold:
\begin{itemize}
    \item An unsupervised methodology for detecting supply chain attacks by leveraging continuous cohesion evaluation.
    \item A novel use of PLMs for function-level cohesion estimation.
    \item Targeted identification of software components requiring cohesion improvement.
\end{itemize}

The rest of this article is structured as follows: 
Section~\ref{sec:related-work} summarizes the prior art related to supply chain attacks, cohesion metrics, and language models for source code analysis.  
The analysis methods, including the data and cohesion change tracking, are elaborated in Section~\ref{sec:method}.
The main research questions and results are discussed in the experimental Section~\ref{sec:experiments}. 
Section~\ref{sec:conclusion} concludes the paper. 

\section{Related Work}
\label{sec:related-work}

\subsection{Supply chain attack}
The increasing development of open-source software has opened another way for attackers to harm software systems. 
Attackers leverage the use of open source during software development, which is enhanced by dependency managers that update, download, and install open-source packages automatically.
This makes it easier for attackers to inject malicious code into the dependency tree of a widely used package, which can help them reach a large number of systems in a short period of time \cite{ohm2020backstabber}.

Attackers attempting to inject malicious code into existing packages may adopt various tactics.
They may mimic legitimate contributors to open-source projects, submitting pull requests with seemingly benign code changes \cite{duan2020towards}. 
Alternatively, attackers may compromise project repositories directly, committing malicious code with weak or compromised credentials, or by social engineering to become project maintainer \cite{zahan2022weak}.
Malicious packages resulting from such attacks can propagate malware across extensive software systems. 
Vu et al.~\cite{vu2020towards} demonstrated that malicious packages surpassed 100,000 downloads in several instances, illustrating the significant impact these attacks can have.
Real-world instances, like the attack on the widely used Homebrew \cite{holmes2018gained}, SolarWinds' Orion platform \cite{sterle2021solarwinds}, and Kaseya's software \cite{robinson2022new} emphasize the severity of these threats.

Numerous solutions have been suggested to mitigate the risks posed by such attacks. 
Duan et al.\cite{duan2020towards} employed established program analysis techniques, including metadata analysis, to identify packages dependent on known malware or sharing similar authors and release patterns. 
They also utilized static analysis to scrutinize installation logic and flows, and dynamic analysis to detect unusual network communication, file usage, and process behavior. 
Their study revealed 339 previously unidentified malicious packages. 
Another approach by Garrett et al.\cite{garrett2019detecting} proposed an anomaly detection-based method, focusing on security-relevant features during version updates. 
Both studies acknowledge the incompleteness of their solutions, emphasizing their role as foundational building blocks.
Recent approaches have also leveraged transformer models for malicious code detection.
Tsfaty and Fire \cite{tsfaty2022malicious} demonstrated that by clustering multiple versions of a function based on code embeddings using Code2Seq~\cite{alon2018code2seq}, they can flag versions that significantly deviate from their nearest cluster.
The main issue with their approach is that they can only monitor function implementations that are highly used across different projects and are supposed to have similar implementations (e.g. \textit{"run"} or \textit{"get"} functions).
Another study used pre-trained transformers to train classifiers that detect malicious code in JavaScript packages \cite{ohm2024using}. 
While they achieved promising results using the backstabber dataset with 10-fold cross-validation, their evaluation methodology did not account for differences between injection patterns in training and test sets, and was dependent on known malware behavior.
Although many solutions were suggested for identifying code injection, a recent analysis on a large dataset of 2105 packages from the PyPI\footnote{\url{https://pypi.org/}} ecosystem showed that malicious code has proven adept at evading detection from existing tools through the employment of numerous anti-analysis techniques \cite{guo2023empirical}.

The persistent vulnerability of the source code supply chain to attacks is underscored by the lack of an integrated and robust protection mechanism~\cite{Malik2023ProtectionMA}. 
Our proposed method represents a more comprehensive solution compared to existing approaches. 
Unlike methods primarily reliant on detecting known malware based on behavior, our approach stands out by detecting side effects inherent in every instance of malware injection into cohesive code.

\subsection{Cohesion metrics}
Code cohesion is a fundamental concept in software engineering, referring to the degree to which the elements within a module belong together \cite{rosenberg1997software}. 
High cohesion is desirable as it indicates that the elements within the module are closely related and work together to perform a specific task. 
Various metrics have been developed to measure cohesion in object-oriented software, aiming to assess the reusability, efficiency, and complexity of software modules. 
These metrics are essential for evaluating the quality of code and identifying opportunities for restructuring to improve the internal structure of software systems~\cite{Sami2020AVC}.

Several approaches have been proposed to address code cohesion measures. 
The Lack of Cohesion in Methods (LCOM) metric is a cohesion measure for object-oriented software, initially defined by Chidamber and Kemerer \cite{chidamber1994metrics}. 
It assesses the level of cohesion in software modules by measuring the number of method pairs that do not share instance variables. 
However, one issue with the LCOM metric is that it fails to differentiate between possible levels of cohesion ~\cite{marcus2005conceptual}.
Tight Class Cohesion (TCC), proposed by Bieman and Kang~\cite{bieman1995cohesion}, measures direct connections between methods through common instance variables. 
Class Connection Metric (CCM), developed by Wasiq~\cite{wasiq2001measuring}, constructs a method connection graph based on shared attributes and method calls. CCM quantifies the connectedness of this graph to measure cohesion, handling transitive connections between methods.
Loose Class Cohesion (LCC), proposed by Bieman and Kang~\cite{bieman1995cohesion}, considers both direct and indirect connections between methods, taking into account transitivity.

Another approach to measure code cohesion involves the use of refactoring to improve the cohesion of object-oriented software. Refactoring aims to alter the internal structure of the code without changing its external functionality, thereby improving cohesion and reducing coupling in the software system~\cite{Rathee2017RestructuringOO}.
Additionally, novel cohesion metrics, such as the Variable Frequency – Inverse Method Frequency (VF-IMF) metric, have been developed to assess the level of cohesion in modules and group module methods to instill high cohesion~\cite{Sami2020AVC}. 
These metrics offer a compromised solution for building high cohesive modules and differentiate between different levels of cohesion, addressing the limitations of traditional cohesion metrics, such as LCOM.
The closest approach to our work is the work of~\cite{haner2023predicting} which tries to solve the challenge of manually tracing the cohesion value in software code, which is essential for evaluating software maintainability. The study focuses on predicting cohesion values, including LCOM2, TCC, and LCC, using machine learning techniques such as KNN, REPTree, multi-layer perceptron, linear regression, support vector machine, and random forest. The research utilizes two different open-source software projects to create datasets and evaluates the performance of various machine learning algorithms for predicting different cohesion metrics. 

Our work is distinct in harnessing the semantic capabilities of large language models to estimate code cohesion. 
By detecting significant deviations, we can identify components needing additional security auditing and analysis. 

\subsection{Code analysis with language models}
Large Language Models (LLMs) have shown significant potential in natural language understanding and programming code processing tasks. Their ability to comprehend and generate human-like code has sparked research interest in leveraging LLMs for code analysis purposes \cite{wang2023review}. 

LLMs have been evaluated for their capabilities in automating code analysis tasks, including the analysis of obfuscated and malicious code~\cite{fang2023large}. 
The findings indicate that LLMs can serve as valuable tools for automating code analysis, albeit with certain limitations. 
Their research contributes to a deeper understanding of the potential and constraints associated with utilizing LLMs in code analysis, paving the way for enhanced applications in this critical domain.
Recent work assessed CodeBERT and ChatGPT for their efficiency in security-oriented program analysis, addressing tasks like code review and code generation. The study delves into the capabilities of LLMs in solving security-related analytic tasks, providing insights into their potential and limitations in addressing software security challenges~\cite{wang2023effectiveness}.

These use cases demonstrate the diverse applications of LLMs in code analysis, ranging from scientific research and data analysis to automating code analysis tasks and addressing security-oriented program analysis challenges. Since LLMs show promise in enhancing productivity and automating various code-related tasks, we choose to use them in order to estimate code cohesion.

\section{Methods}
\label{sec:method}
\begin{figure}
    \centering
    \includegraphics[width=1\linewidth]{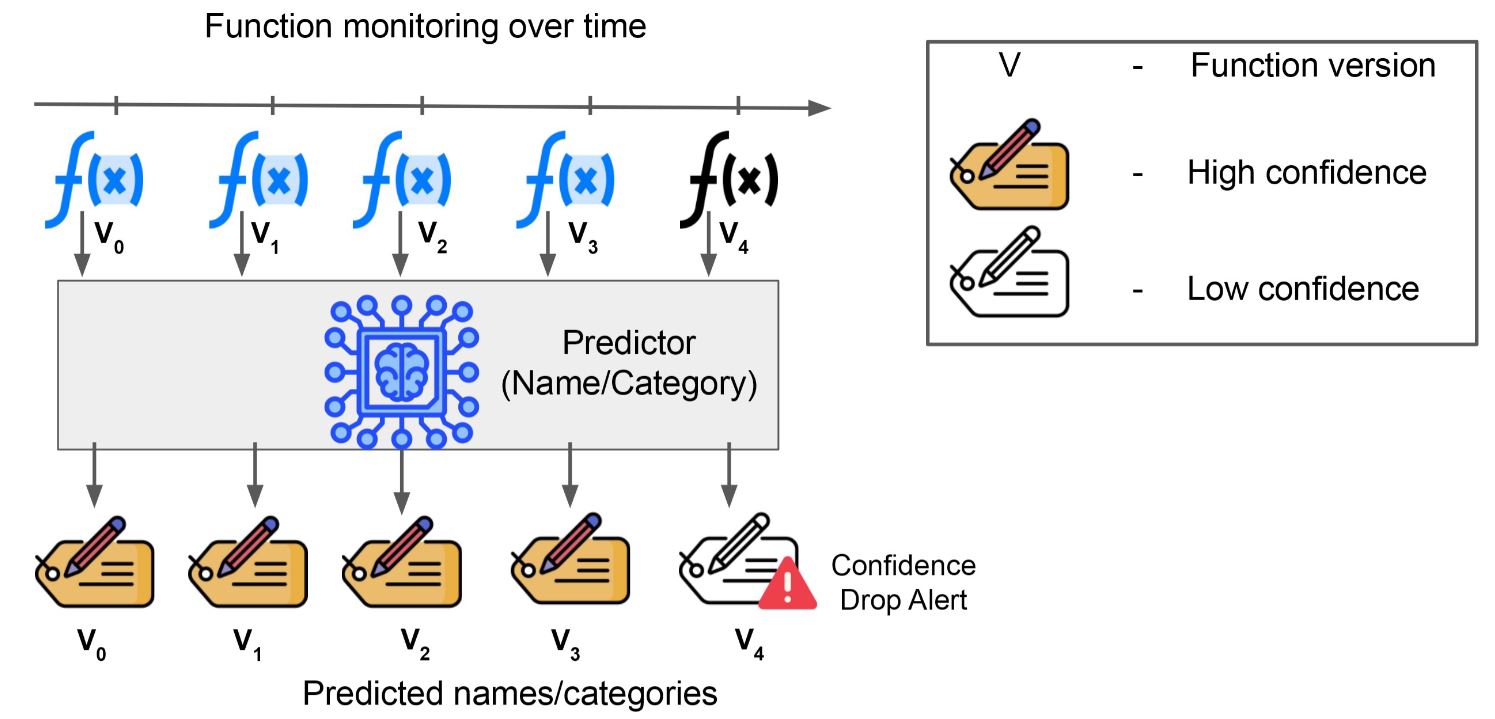}
    \caption{Overview of cohesion monitoring to detect potential security compromises. By analyzing cohesion metrics over successive version releases, we can identify significant decreases to flag for investigation.}
    \label{fig:method-overview}
\end{figure}

In this section, we present our methodology for highlighting spurious code insertions by monitoring for significant drops in code cohesion. 
We begin by describing the dataset used to evaluate our code cohesion metrics over time. 
Next, we demonstrate how fine-tuned language models are used to quantify code cohesion through a function name prediction task. 
Finally, we describe our approach to simulating software compromise by injecting malicious code into functions. 
The primary goal of our methodology is to monitor software over time and detect significant decreases in code cohesion that may indicate a security compromise. 
Figure~\ref{fig:method-overview} illustrates this process by analyzing cohesion metrics for each new version release, we can flag suspicious drops for further investigation.




\subsection{Dataset}
\label{sec:dataset}
We collected a corpus of 109,924 C++ functions from 369 widely used open-source GitHub projects\footnote{A list of the repositories used can be found at \url{https://tinyurl.com/github-repositories}}. To expand this dataset, we gathered all publicly available versions of these functions from their complete Git histories.

A function version represents either the initial implementation or any subsequent modification of a function due to a commit. 
To ensure data quality and completeness, the collection process involved the following steps:
\begin{enumerate}
    \item Gathering repository metadata and commit histories.
    \item Filtering commits to retain only those with file modifications.
    \item Downloading the modified files.
    \item Extracting functions using Clang\footnote{\url{https://clang.llvm.org/}}.
\end{enumerate}

Each function was assigned a unique identifier combining its name, argument types, and file location, allowing precise tracking across versions even when implementations evolved. Using these identifiers, we eliminated unchanged duplicates while maintaining comprehensive version histories. 
After filtering out functions with only a single version, the final dataset comprised 479,996 versions of 54,706 unique functions, with an average of 8.7 and a median of 3 versions per function.

For analysis, we focused on functions with multiple versions, yielding 425,290 consecutive version pairs. These pairs capture natural changes resulting from regular maintenance and feature development, forming a robust foundation for examining cohesion dynamics over time and identifying deviations caused by code injections.
Detailed statistics for the full dataset are presented in Table~\ref{tab:dataset-summary}.

\begin{table}[h!]
\caption{Dataset Summary Statistics}
\label{tab:dataset-summary}
\centering
\begin{tabular}{|c|l|c|}
\hline
& \textbf{Statistic} & \textbf{Value} \\ \hline
\multirow{10}{0.3cm}{\rotatebox{90}{Open source C++}} & \#Projects & 369 \\ \cline{2-3}
& Avg. lines of code per project & 131,758 \\ \cline{2-3}
& Linux-compatible projects & 369 (100\%) \\ \cline{2-3}
& Windows-compatible projects & 115 (31.25\%) \\ \cline{2-3}
& \#Function versions & 479,996 \\ \cline{2-3}
& \#Unique functions & 54,706 \\ \cline{2-3}
& Average versions per function & 8.7 \\ \cline{2-3}
& Median versions per function & 3 \\ \cline{2-3}
& Avg. lines of code per function  & 14.65 \\ \cline{2-3}
& Avg. tokens per function & 197.58 \\ \hline \hline
\multirow{4}{0.3cm}{\rotatebox{90}{Malware}} & \#Malicious code examples& 9 \\ \cline{2-3}
& Avg. Lines of Code per malware & 6.44 \\ \cline{2-3}
& Avg. tokens per malware & 77.66 \\ \cline{2-3}
& \#Consecutive version pairs & 425,290 \\ \hline
\end{tabular}

\end{table}


\subsection{Function name prediction}
\label{sec:name-prediction-method}
The function name prediction approach leverages a pre-trained language model to predict an appropriate name for a function based solely on its body text. 
Appropriately named functions typically exhibit high cohesion, as the name succinctly captures their singular, focused purpose. 
Intuitively, a coherent function performing a single focused task should be easier for the model to name accurately compared to one with disjoint or unrelated logic.
To operationalize this idea, we employ a systematic methodology based on masked language modeling (MLM).
Using this technique, we replace the function name with a sequence of token masks (e.g., \texttt{<mask1>}, \texttt{<mask2>}), and the model predicts the most probable terms to fill these masks, effectively generating a name for the function.

For a given function $f$, we perform the fill-mask operation on its name using token counts ranging from one to eight, which covers the typical range of function names in practice.
The confidence for $f$ when using $n$ tokens is calculated using the harmonic mean of the token probabilities, as shown in Equation~\ref{eq:confidence}.

\begin{equation} 
\label{eq:confidence}
\textit{Confidence}(f, n) = \frac{n}{sum_{i=1}^n{{p_i}^{-1}}}
\end{equation}

Here, \(p_i\) is the probability of the \(i\)-th token. 
The harmonic mean penalizes low-probability tokens, ensuring that the metric reflects consistent confidence across all predicted tokens.
Then, we compute the confidence for each token count and select the maximum confidence as the cohesion metric, defined in Equation~\ref{eq:max-confidence}.

\begin{equation} 
\label{eq:max-confidence}
\textit{NPC}(f) = \max(\textit{Confidence}(f, n) \mid n \in \{1, \ldots, 8\})
\end{equation}

This metric, referred to as the name prediction cohesion (NPC), provides an aggregate measure of how well the model can generate a cohesive, meaningful name for the function.
We also refer to the token count that yields the highest confidence as the optimal token count (OTC), which indicates the number of tokens that best capture the semantic intent of the function.
Using this terminology, we can also define NPC of function $f$ as follows:
\begin{equation} 
\textit{NPC}(f) = \textit{Confidence}(f, \textit{OTC}(f))
\end{equation}

This approach not only quantifies the alignment between function names and their logic but also provides a systematic method for evaluating cohesion across varying token counts. 
To implement this methodology, we used the CodeBERTCpp model, which is pre-trained and fine-tuned on C++ code, making it well-suited for analyzing the syntax and semantics of functions in this language.

\subsection{Code Injection} 
\label{sec:code-injection}
To evaluate the sensitivity of our cohesion quantification techniques to compromised integrity, we inject malicious code segments into functions and analyze their impact on cohesion assessments. 
The malicious code corpus consists of segments exhibiting common behaviors such as exfiltrating sensitive data, performing privilege escalation, and enabling remote code execution. 
These segments are sourced from publicly available repositories of known malware code\footnote{The malware code used in this study can be accessed at \url{https://tinyurl.com/malicious-implants}}.
After selecting these segments, we inject them at three strategic locations within functions: the beginning (after the declaration but before existing instructions), middle (after 50\% of the original lines), or end (after all existing code). 
We ensure the modified code maintains syntactic validity.

\begin{figure*}[t]
    \centering
    \begin{subfigure}[t]{0.48\linewidth}
        \raisebox{8.7mm}{\includegraphics[width=\linewidth]{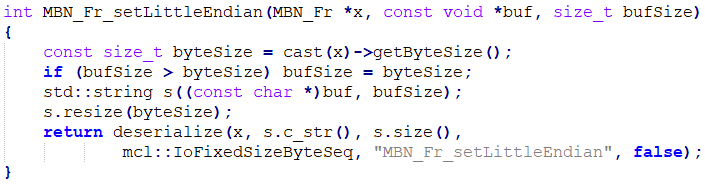}}
        \caption{Function before malware injection.}
        \label{fig:before-injection}        
    \end{subfigure}
    \hfill
    \begin{subfigure}[t]{0.48\linewidth}
        \includegraphics[width=\linewidth]{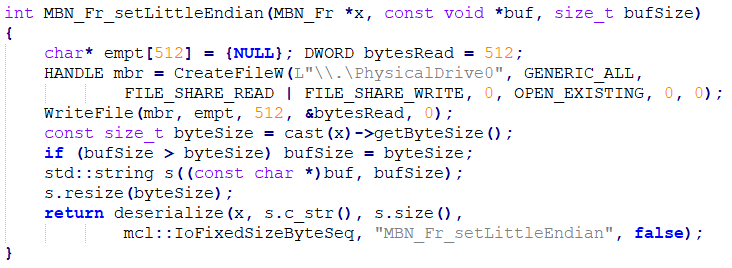}
        \caption{Function after malware injection at the beginning of a function.}
        \label{fig:after-injection}
    \end{subfigure}
    \caption{Example of malware injection into a function.}
    \label{fig:injection-example}
\end{figure*}

In Figure~\ref{fig:injection-example}, we show an example of a function before and after injection. The injected payload writes directly to the master boot record (MBR), demonstrating how we simulate malware injection.

\subsection{Cohesion Change Quantification}
To evaluate the cohesion change after code injection or between consecutive versions of the same function, we use two complementary methods: The \textbf{Cohesion Difference (CD)} is defined as the NPC difference between functions $f_1$ and $f_2$, detailed in Equation~\ref{eq:cd}.

\begin{equation}
    \textit{CD}(f_1, f_2) = \textit{NPC}(f_1) - \textit{NPC}(f_2)
    \label{eq:cd}
\end{equation}

The \textbf{Optimal Token Count Difference (OTCD)} is the confidence  difference of $f_1$ optimal token count (OTC), detailed in Equation~\ref{eq:otcd}:
\begin{equation}
    \textit{OTCD}(f_1, f_2) = \textit{NPC}(f_1) - \textit{Confidence}(f_2, \textit{OTC}(f_1))
    \label{eq:otcd}
\end{equation}

These methods provide complementary insights into cohesion changes. 
The CD captures the overall cohesion drift, reflecting shifts in the model's maximum confidence about the function's coherence, regardless of token count.
The OTCD captures disruptions specific to the original function's optimal representation, highlighting how well the modified function maintains the semantic intent of the original token count.
Together, they help quantify both structural and semantic disruptions caused by code injection or iterative changes between function versions.
A positive value in either metric indicates a decrease in cohesion, while negative values suggest increased cohesion.

\section{Experiments}
\label{sec:experiments}
To evaluate the effectiveness of cohesion-based security monitoring, we investigate four key research questions (RQ):
\begin{RQ}
How does malware injection affect cohesion when measured through function name prediction?
    \label{rq:injection-impact-cohesion-1a}
\end{RQ}
\begin{RQ}
What is the distribution of cohesion across functions in real-world software?
    \label{rq:cohesion-distribution}
\end{RQ}
\begin{RQ}
How effectively can cohesion metrics distinguish between legitimate maintenance changes and malicious code injections?
    \label{rq:legitima-injected-auc}
\end{RQ}

\subsection{\textbf{RQ\ref{rq:injection-impact-cohesion-1a} -} How does malware injection impact cohesion when measured using function name prediction?} 

In well-structured code, functions should have clear and descriptive names that convey their purpose and functionality, helping developers understand what the function does without needing to examine its implementation details. 
Names that are too short may be ambiguous and insufficiently descriptive, while overly long names can become cumbersome and harder to read.
When code injection occurs, it often introduces functionality that diverges from the function's original purpose, creating a semantic gap between the function's name and its actual behavior.
This misalignment between name and implementation can serve as a potential indicator of malicious modifications.

To investigate this, we explore three hypotheses (H) related to function name prediction:
\begin{hypothesis}
For every function, there is an optimal number of tokens comprising its name.
    \label{en:name-1}
\end{hypothesis}
\begin{hypothesis}
Code injection changes the optimal number of tokens in a function's name.
    \label{en:name-2}
\end{hypothesis}
\begin{hypothesis}
Code injection reduces the function cohesion.
    \label{en:name-3}
\end{hypothesis}

\begin{figure*}
    \centering
    \includegraphics[width=0.8\linewidth]{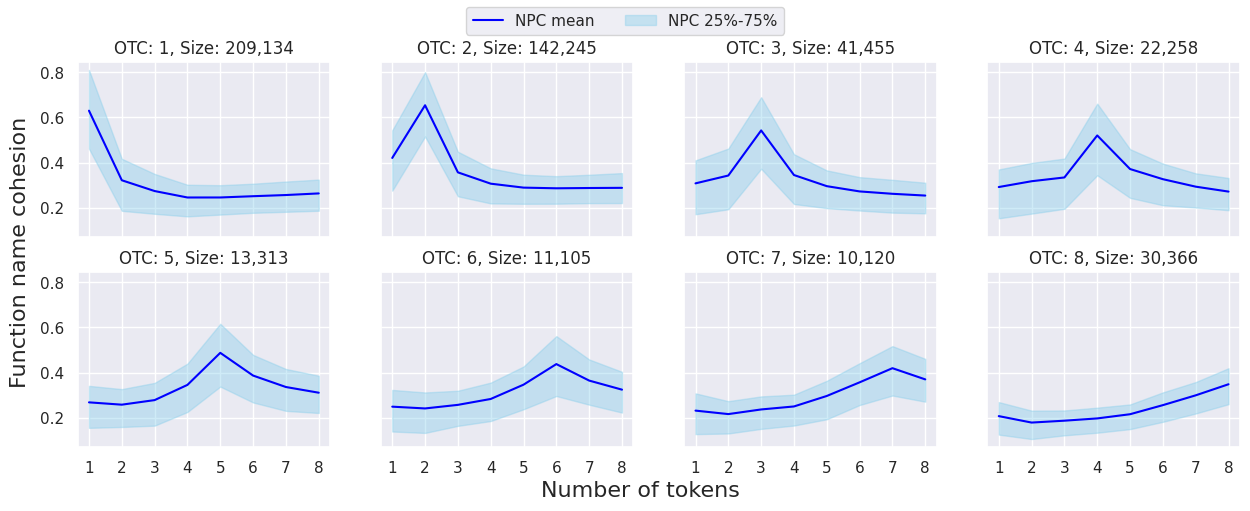}
    \caption{Distribution of name prediction confidence across different token counts. Each subplot represents functions grouped by their optimal number of tokens. The solid line shows the mean confidence, while the shaded area represents the 25th to 75th percentile range. The x-axis shows the number of tokens, and the y-axis represents the prediction confidence score.}
    \label{fig:optimal-tokens}
\end{figure*}

To test these hypotheses, we analyzed 479,996 functions from the dataset, predicting names ranging from one to eight tokens.
We grouped functions based on their optimal token count (OTC) and displayed the aggregated mean and quantiles of the results for maximum confidence, as shown in Figure~\ref{fig:optimal-tokens}.

The analysis reveals distinct peaks in confidence—both in mean and quantiles—corresponding to the OTCs. 
These peaks demonstrate that functions naturally gravitate toward specific name lengths that optimize their descriptiveness and predictability, providing quantitative support for H\ref{en:name-1}.

\begin{figure*}
    \centering
    \includegraphics[width=0.8\linewidth]{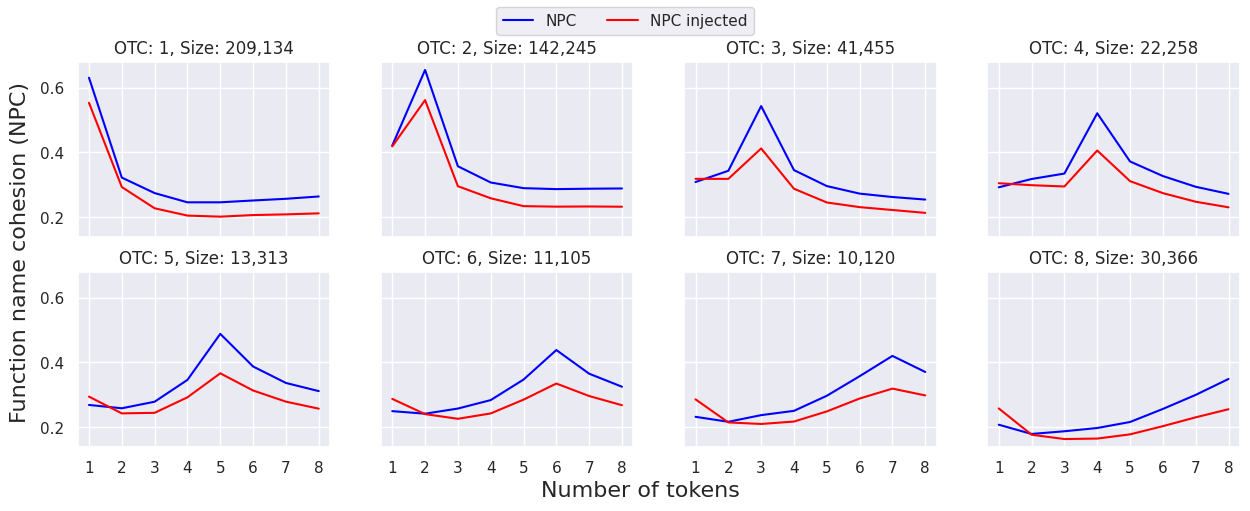}
    \caption{Impact of code injection on name prediction confidence. Each subplot shows functions grouped by their original optimal token count. Blue lines represent the original code's confidence scores, while red lines show scores after injection. The x-axis represents the number of tokens, and the y-axis shows the prediction confidence score.}
    \label{fig:optimal-tokens-with-injections}
\end{figure*}

Next, we injected malware (see Section~\ref{sec:code-injection}) at the beginning, middle, and end of each function and applied the same process. 
The results are presented in Figure~\ref{fig:optimal-tokens-with-injections}.
After code injection, the OTC shifts to one token for many functions, and the overall maximum confidence decreases. 
This simplification of function names after injection supports H\ref{en:name-2}, showing that external code alterations negatively impact natural naming conventions. 
Furthermore, the drop in confidence post-injection supports H\ref{en:name-3}, indicating that code injection reduces the predictability of function names.

Following this, we conducted a comprehensive evaluation of cohesion changes under two scenarios: malicious code injection and natural code evolution. 
For the injection scenario, we modified each function by inserting different types of malicious code at three locations (beginning, middle, and end) and averaged the results (see Section~\ref{sec:code-injection}).
We quantified these changes using two metrics: the cohesion difference (CD) and optimal token count difference (OTCD), measuring the changes in these values before and after injection.
We also performed a separate analysis on high-cohesion functions (NPC > 0.5) to determine whether injection effects were more pronounced in well-structured code.
Since high-cohesion functions have a more predictable structure and naming consistency, we hypothesize that malicious injections will cause greater disruptions, making them easier to detect.

To establish a baseline for comparison, we also examined how cohesion naturally evolves across multiple versions of the same function. Using the same metrics (CD and OTCD), we analyzed 425,290 consecutive version pairs, measuring how cohesion changes during normal development. 
Detailed results on the impact of code injection on cohesion metrics across three injection positions and between consecutive version (baseline) pairs are presented in Table~\ref{tab:injection-impact}.

\begin{table}[t]
\caption{Impact of code injection on cohesion metrics}
\label{tab:injection-impact}


\begin{tabular}{l|l|ll}
 & \textbf{Position} & \textbf{CD} & \textbf{OTCD}  \\
\midrule
\multirow{4}{*}{\textbf{\begin{tabular}[c]{@{}l@{}}All \\ functions\end{tabular}}} & Beginning & 0.027$\pm$0.224* & 0.105$\pm$0.227* \\
 & Mid & 0.040$\pm$0.173* & 0.086$\pm$0.178* \\
 & End & 0.038$\pm$0.172* & 0.083$\pm$0.177* \\
 & Baseline & 0.0005$\pm$0.12 & 0.029$\pm$0.131 \\
 \midrule
\multirow{4}{*}{\textbf{\begin{tabular}[c]{@{}l@{}}High \\ cohesion \\ (\textgreater{}0.5)\end{tabular}}} & Beginning & 0.081$\pm$0.207* & 0.142$\pm$0.231* \\
 & Mid & 0.077$\pm$0.163* & 0.11$\pm$0.183* \\
 & End & 0.074$\pm$0.161* & 0.106$\pm$0.18* \\
 & Baseline & 0.021$\pm$0.112 & 0.045$\pm$0.138 \\
\end{tabular}
\\ \\
* indicates statistical significance (p < 0.05) compared to the consecutive versions (baseline).
Values are presented as mean $\pm$ standard deviation.
\end{table}

Our results demonstrate that code injection significantly impacts function cohesion across all injection positions, with CD values ranging from 0.027 to 0.038 and OTCD values ranging from 0.083 to 0.105.
Notably, the changes in cohesion between consecutive versions of the same function (mean CD = 0.0005, mean OTCD = 0.031) are significantly (p < 0.05) smaller than those observed after code injection (minimum CD = 0.027, minimum OTCD = 0.083), with differences of over an order of magnitude in CD values.
For high-cohesion functions, the impact of injection was substantially stronger, with CD values ranging from 0.074 to 0.081 and OTCD values ranging from 0.106 to 0.142. 
The increased sensitivity in high-cohesion functions is particularly notable close to the beginning, where CD increased to 0.081 (200\% higher than the general case) and OTCD reached 0.142 (35\% higher). 
This suggests that well-structured code is indeed more sensitive to malicious modifications, making injection detection more reliable in high-quality codebases.

These findings provide strong support for our hypotheses: functions exhibit an optimal token count (H\ref{en:name-1}), which is disrupted by injection (H\ref{en:name-2}). Also, both CD and OTCD metrics detect these disruptions showing statistically significant reductions in cohesion after injection (H\ref{en:name-3}). The clear distinction between natural evolution and injected modifications, particularly in CD measurements, demonstrates the effectiveness of our name prediction cohesion (NPC) metric for detecting potential malicious code insertions.


\subsection{\textbf{How is cohesion distributed across functions, and how is it affected by maintenance changes and function size?}}

We examined the distribution and characteristics of function cohesion from three perspectives: overall distribution, maintenance impact, and the relationship with function size.

\begin{figure}
    \centering
    \includegraphics[width=1\linewidth]{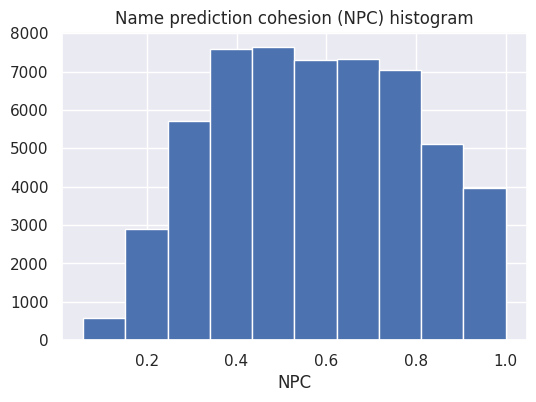}
    \caption{Distribution of name prediction cohesion (NPC) scores across functions.}
\label{fig:cohesion-histogram}
\end{figure}

To understand how cohesion naturally varies across different functions, we first computed a cohesion histogram for all consecutive function version pairs in the dataset using the name prediction cohesion (NPC) metric. 
The results are presented in Figure \ref{fig:cohesion-histogram}.
The histogram reveals that the majority of functions (approximately 63\%) exhibit high cohesion (greater than 0.5).
This aligns with software engineering principles and expectations, as these functions come from widely used open-source packages that typically undergo code reviews and maintain quality standards.

To understand how cohesion changes during normal maintenance vary across different cohesion levels, we examined the relationship between a function's base cohesion level and its CD and OTCD measurements.
We calculated the Pearson correlation between the function's NPC score and both its CD and OTCD values during normal maintenance changes.
The analysis revealed significant (p-value < 0.05) positive Pearson's correlations: 0.951 for CD and 0.894 for OTCD with respect to cohesion levels.
These strong correlations indicate that functions with higher cohesion tend to experience larger cohesion drops during maintenance. 
This can be explained mathematically, due to the fact that larger values of NPC (range between 0 to 1) have more "room" to fall toward 0 than lower values.

\begin{figure}
    \centering
    \includegraphics[width=1\linewidth]{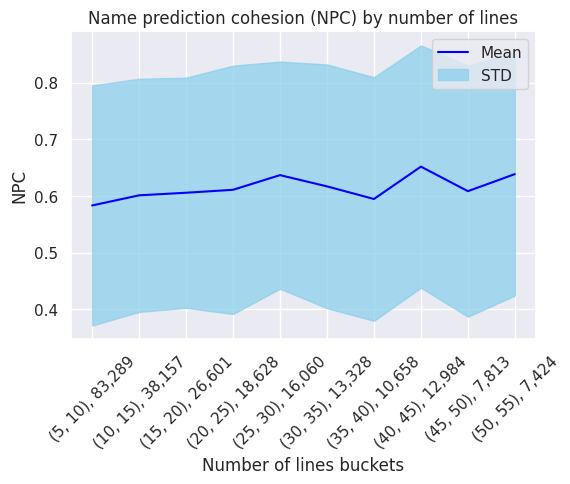}
    \caption{Mean and standard deviation of name prediction cohesion (NPC) across function sizes. Each bucket represents a five-line interval, showing consistent cohesion levels regardless of function length.}
\label{fig:Name-prediction-based-cohesion-distribution-for-legitimate-functions}

\end{figure}

Next, we explored the relationship between cohesion scores and the number of lines of code. 
To achieve this, we grouped functions into buckets based on their size, with each bucket representing a five-line interval (see Figure \ref{fig:Name-prediction-based-cohesion-distribution-for-legitimate-functions}).

The analysis revealed that the average and standard deviation of code cohesion remained consistent across different function sizes (about $0.608 \pm 0.212$) and found no significant correlation between NPC and function size (p-value = 0.186).
This indicates that the NPC metric provides a reliable measure of function cohesion, independent of function size.

\begin{figure}
    \centering
    \includegraphics[width=1\linewidth]{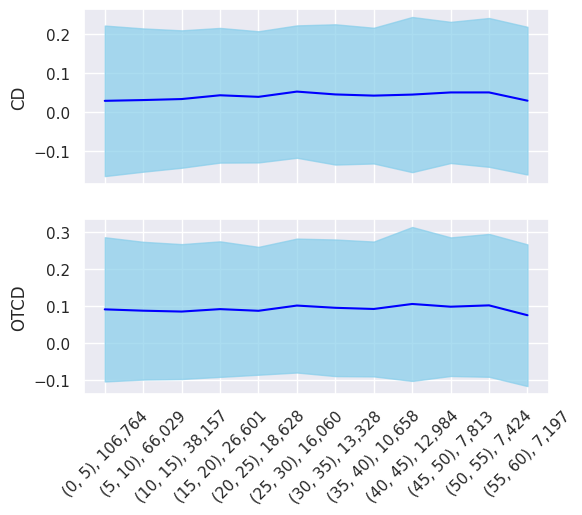}
    \caption{Impact of code injection on cohesion across different function sizes. Each bucket represents a five-line interval, showing the mean cohesion difference (CD and OTCD) and standard deviation after malware injection. Function size does not affect CD and OTCD.}
    \label{fig:Cohesion difference (CD) by number of lines}
\end{figure}

To further investigate how the relative size of injected code compared to function size impacts cohesion, we calculated the mean cohesion difference (CD) and optimal token count difference (OTCD) for malicious pairs in each bucket. 
Across all buckets, the mean cohesion difference ranged from 0.028 to 0.051, with a standard deviation between 0.167 and 0.198 (see Fig.\ref{fig:Cohesion difference (CD) by number of lines}).

\begin{table*}
\centering
\caption{Adjusted P@100 scores for various malicious-to-benign ratios using different cohesion metrics.}
\begin{tabular}{lr|llll|c}
 & \textbf{Ratio} & \multicolumn{1}{l}{\textbf{CD}} & \multicolumn{1}{l}{\textbf{OTCD}} & \multicolumn{1}{l}{\textbf{CDz}} & \multicolumn{1}{l|}{\textbf{OTCDz}} & \multicolumn{1}{l}{\textbf{\#Malwares}} \\ \hline
\multirow{3}{*}{\begin{tabular}[c]{@{}l@{}}All \\ function\end{tabular}} & 1:100 & 1.41\% & 1.49\% & 2.65\% & 3.41\% & 4799 \\
 & 1:1,000 & 0.13\% & 0.17\% & 0.27\% & 0.35\% & 479 \\
 & 1:10,000 & 0.04\% & 0.03\% & 0.06\% & 0.08\% & 47 \\ \hline
\multirow{3}{*}{\begin{tabular}[c]{@{}l@{}}High \\ cohesion\end{tabular}} & 1:100 & 82.21\% & 4.53\% & 100.00\% & 100.00\% & 3030 \\
 & 1:1,000 & 8.29\% & 0.42\% & 36.41\% & 22.30\% & 303 \\
 & 1:10,000 & 2.97\% & 0.14\% & 12.47\% & 7.82\% & 30
\end{tabular}
\label{tab:detection-results}
\end{table*}

The results show no significant correlation between function size and either CD (p = 0.134) or OTCD (p = 0.562).
This provides additional evidence that function size does not affect our function cohesion metric, making it a robust measure for detecting code injection across functions of varying lengths.



\subsection{\textbf{RQ\ref{rq:legitima-injected-auc} -}How sensitive are our code cohesion metrics in distinguishing between legitimate and injected code changes?}

In this experiment, we evaluate the effectiveness of our code cohesion metrics in distinguishing between normal maintenance changes and malicious code injections. 
We analyzed 425,290 function pairs across 369 publicly available repositories to simulate real-world detection scenarios.

To reflect realistic attack scenarios where malicious code represents a small fraction of the codebase, we simulated three malicious-to-benign ratios: 1:100, 1:1,000, and 1:10,000. 
In each ratio, the functions were chosen randomly (based on the ratio) of being injected with malware at a random location (beginning, middle, or end). 
Next, we calculate CD and OTCD scores between consecutive versions or pre/post-injection pairs.
Finally, we evaluate detection effectiveness using precision at 100 (P@100), which measures detection accuracy in the top 100 ranked functions.
We conducted these simulations 1000 times for each ratio and averaged the results.
For simulations with fewer than 100 injected functions (1:10,000 ratio), we adjust P@100 by dividing by the maximum possible precision.

To account for the dependency between cohesion differences and baseline cohesion scores, we introduced Standardized Cohesion Difference (CDz) and Standardized Optimal Token Count Difference (OTCDz).
These metrics represent the number of standard deviations from the mean difference for functions with similar cohesion scores.
The CDz and OTCDz are calculated by grouping functions into 20 buckets of 0.05 cohesion width (based on their NPC), computing the mean and standard deviation for each bucket, and standardizing differences relative to their bucket statistics.
For instance, consider a function that initially has a cohesion score of 0.73 (version $v_1$) and drops to 0.62 (version $v_2$), resulting in a CD of 0.11. To compute CDz, we first determine the mean and standard deviation of CD within the corresponding $[0.7,0.75]$ cohesion bucket ($\mu=0.057,\sigma=0.054$). The standardized difference is then calculated as $CDz = (0.11-0.057)/0.054=0.981$, indicating the magnitude of deviation relative to similar functions.

Our results indicate that cohesion-based metrics effectively distinguish between legitimate and malicious code changes, particularly in high-cohesion functions. 
At a 1:100 ratio, detection performance in high-cohesion functions was significantly higher (82.21\%) compared to the overall detection rate (1.41\%), with standardized metrics (CDz, OTCDz) achieving perfect detection (100\%). 
As the injection ratio became more extreme (1:10,000), detection performance decreased but remained meaningful, with CDz and OTCDz achieving 12.47\% and 7.82\% detection rates in high-cohesion functions—representing a \textbf{three-order-of-magnitude reduction in manual review effort}. 
Furthermore, \textbf{standardized metrics consistently outperformed non-standardized ones}, highlighting the importance of adjusting for baseline cohesion levels.
CD-based metrics generally provided stronger detection capabilities than OTCD-based ones, reinforcing their effectiveness in identifying anomalous code changes. 
These findings demonstrate that cohesion-based anomaly detection, particularly when leveraging standardized metrics and high-cohesion functions, is a powerful approach for detecting malware injections even in highly imbalanced real-world scenarios.

\section{Conclusion}
\label{sec:conclusion}
This study presents an unsupervised approach for detecting supply chain attacks by analyzing cohesion disruptions in source code. Our findings demonstrate that name-prediction-based cohesion metrics can effectively capture the impact of malicious code injections by identifying deviations in function cohesion patterns.

We first examined how malware injection affects function cohesion and found that injected code disrupts natural naming patterns, reducing the confidence of the name prediction model. This confirms the sensitivity of our metric to malicious modifications. Additionally, our analysis showed that cohesion scores remain stable across function sizes, reinforcing the metric’s reliability. However, we found that cohesion drop metrics (CD and OTCD) are influenced by a function’s initial cohesion level, indicating that normalization (CDz and OTCDz) improves their robustness.

To assess real-world applicability, we simulated various benign-to-malicious ratios, demonstrating that monitoring high-cohesion functions with NPC effectively detects injected functions, achieving a P@100 of 36.41\% at a 1:1,000 ratio and 12.47\% at 1:10,000. These results highlight the potential of cohesion metrics as a lightweight, scalable solution for assisting in the identification of supply chain attacks.

Despite these promising results, our approach has limitations. Our injection strategy relies on controlled placements, whereas real-world attacks may use more sophisticated techniques to evade detection. Additionally, while our analysis of 369 repositories did not uncover concrete evidence of supply chain attacks, refining our method with broader datasets and real-world malware samples could enhance detection capabilities.
Future research could enhance our approach by incorporating more advanced language models for cohesion evaluation and extending the analysis to additional programming languages. 
Overall, our findings suggest that name-prediction-based cohesion metrics offer a viable, automated solution for detecting code injections, contributing to the broader goal of securing source code supply chains.

\bibliographystyle{unsrt}  
\bibliography{references}  
\end{document}